\documentclass{my-ijgmmp}

\usepackage{hyperref}

\usepackage[noadjust]{cite}

\usepackage{amssymb} 
\usepackage{amsfonts} 
\usepackage{amsmath} 
\usepackage{slashed} 

\usepackage{egothic}%
\usepackage{pgothic}%
\usepackage{yfonts}%
 \usepackage{yhmath} 
  
 \usepackage{yhmath} 

\DeclareMathOperator{\Tr}{Tr} 

\begin{document} 


%
\catchline{}{}{}{}{}
%

\title{\Large The De Donder--Weyl 
Hamiltonian formulation of TEGR \\ \vspace{4pt}  
and its quantization}

\author{\large I. V. Kanatchikov}

\address{\normalsize 
National Quantum Information Centre KCIK,  
Gdansk, Poland \\
IAS-Archimedes Project, C\^{o}te d’Azur, France \\
}



\maketitle

\begin{history}
\end{history}

\begin{abstract}  \small  
The De Donder-Weyl (DW) Hamiltonian theory of fields treats space and time variables on equal footing. Its quantization, called precanonical quantization, leads to a hypercomplex generalization of quantum formalism to field theory as it follows from the quantization of Poisson-Gerstenhaber brackets defined on differential forms. Our recent work on precanonical quantization of general relativity is extended to the teleparallel equivalent of general relativity (TEGR) in tetrad Palatini formulation. The covariant precanonical Schr\"odinger equation for quantum TEGR and the relevant operators are constructed from the quantization of generalized Dirac brackets calculated using the constraints analysis generalized to the DW Hamiltonian theory. Our analysis of the ordering ambiguities in the precanonical Schr\"odinger equation allows us to estimate the contribution to the cosmological constant from the quantum TEGR and argue its consistency with the observed value, albeit with the current error of estimation of 13 orders of magnitude due to the theoretical uncertainties in the relation between the scale $\varkappa$ introduced by precanonical quantization and the mass gap in the pure gauge sector of QCD. 
\end{abstract}

\vspace{20pt}

\newcommand{\cl}{C \kern -0.1em \ell}  

\newcommand{\JJB}[1]{\JJ_{#1}}
\newcommand{\LLB}[1]{\LL_{#1}}

\newcommand{\w}{\wedge}
\newcommand{\bigw}{\bigwedge}
\newcommand{\dw}{\mathbin{\dot\wedge}}
\newcommand{\dwedge}{\mathbin{\dot\wedge}}

\newcommand{\by}{\mathbf{y}}
\newcommand{\ba}{\mathbf{a}}
\newcommand{\bb}{\mathbf{b}}

\newcommand{\BF}{\mathbb{F}}
\newcommand{\BZ}{\mathbb{Z}}
\newcommand{\BR}{\mathbb{R}}
\newcommand{\BC}{\mathbb{C}}


\newcommand{\beq}{\begin{equation}}
\newcommand{\eeq}{\end{equation}}
\newcommand{\beqa}{\begin{eqnarray}}
\newcommand{\eeqa}{\end{eqnarray}}
\newcommand{\nn}{\nonumber}
\newcommand{\half}{\frac{1}{2}}
\newcommand{\xt}{\tilde{X}}
\newcommand{\com}[2]{[#1,#2]_{-}} 
\newcommand{\acom}[2]{[#1,#2]_{+}} 
\newcommand{\compm}[2]{[#1,#2]_{\pm}}

\newcommand{\lie}[1]{\pounds_{#1}}
\newcommand{\co}{\circ}
\newcommand{\sgn}[1]{(-1)^{#1}}
\newcommand{\lbr}[2]{ [ \hspace*{-1.5pt} [ #1 , #2 ] \hspace*{-1.5pt} ] }
\newcommand{\lbrpm}[2]{ [ \hspace*{-1.5pt} [ #1 , #2 ] \hspace*{-1.5pt}
 ]_{\pm} }
\newcommand{\lbrp}[2]{ [ \hspace*{-1.5pt} [ #1 , #2 ] \hspace*{-1.5pt} ]_+ }
\newcommand{\lbrm}[2]{ [ \hspace*{-1.5pt} [ #1 , #2 ] \hspace*{-1.5pt} ]_- }


\newcommand{\pbr}[2]{ \{ \hspace*{-2.6pt} [ #1 , #2\hspace*{1.4 pt} ] 
\hspace*{-2.6pt} \} }
\newcommand{\nbr}[2]{ [ \hspace*{-1.5pt} [ #1 , #2 \hspace*{0.pt} ] 
\hspace*{-1.3pt} ] }

\newcommand{\we}{\wedge}
\newcommand{\nbrpq}[2]{\nbr{\xxi{#1}{1}}{\xxi{#2}{2}}}
\newcommand{\lieni}[2]{$\pounds$${}_{\stackrel{#1}{X}_{#2}}$  }

\newcommand{\rbox}[2]{\raisebox{#1}{#2}}
\newcommand{\xx}[1]{\raisebox{1pt}{$\stackrel{#1}{X}$}}
\newcommand{\xxi}[2]{\raisebox{1pt}{$\stackrel{#1}{X}$$_{#2}$}}
\newcommand{\ff}[1]{\raisebox{1pt}{$\stackrel{#1}{F}$}}
\newcommand{\dd}[1]{\raisebox{1pt}{$\stackrel{#1}{D}$}}
\newcommand{\der}{\partial}

\newcommand{\Lm}{\bigwedge^*}

\newcommand{\inn}{\hspace*{2pt}\raisebox{-1pt}{\rule{6pt}{.3pt}\hspace*
{0pt}\rule{.3pt}{8pt}\hspace*{3pt}}}
\newcommand{\sro}{Schr\"{o}dinger\ }
\newcommand{\vol}{\omega}
               \newcommand{\dvol}[1]{\der_{#1}\inn \vol}

\newcommand{\bd}{\mbox{\bf d}}
\newcommand{\bder}{\mbox{\bm $\der$}}
\newcommand{\bI}{\mbox{\bm $I$}}

\newcommand{\be}{\beta} 
\newcommand{\gab}{\bar{\gamma}} 
\newcommand{\ga}{\gamma} 
\newcommand{\de}{\delta} 
\newcommand{\gmu}{\bar{\gamma}^\mu}
\newcommand{\gnu}{\bar{\gamma}^\nu}
\newcommand{\ka}{\varkappa}
\newcommand{\hka}{\hbar\varkappa}

\newcommand{\lapl}{\bigtriangleup}
\newcommand{\psib}{\overline{\psi}}
\newcommand{\Psib}{\overline{\Psi}}
\newcommand{\Phib}{\overline{\Phi}}
\newcommand{\derts}{\stackrel{\leftrightarrow}{\der}}
\newcommand{\what}[1]{\widehat{#1}}
\newcommand{\deltab}{\overline{\delta}}

\newcommand{\pib}{\overline{\pi}}
\newcommand{\Cb}{\overline{C}}
\newcommand{\ub}{\overline{u}}

\newcommand{\bx}{{\mathbf{x}}}
\newcommand{\bk}{{\bf k}}
\newcommand{\bq}{{\bf q}}
\newcommand{\bee}{{\bf e}}

\newcommand{\omk}{\omega_{\bf k}} 
\newcommand{\lpl}{\ell}
\newcommand{\zb}{\overline{z}} 

\newcommand{\dv}{\mbox{\sf d}}

\newcommand{\BPsi}{{\bf \Psi}} 
   

\newcommand{\BPhi}{{\bf \Phi}} 
\newcommand{\BXi}{{\bf \Xi}} 
\newcommand{\BH}{{\bf H}} 
\newcommand{\BS}{{\bf S}} 
\newcommand{\BN}{{\bf N}}

\newcommand{\hy}{\hat{y}}
\newcommand{\hP}{\hat{P}}
\newcommand{\hH}{\what{H}}

\newcommand{\myvdots}{\mbox{\raisebox{-2pt}{$\vdots$}}}

\newcommand{\fbar}{\bar{f}}

\newcommand{\rd}{\mathrm{d}} 
\newcommand{\re}{\mathrm{e}} 
\newcommand{\ri}{\mathrm{i}} 
\def\semicolon{\nobreak\mskip2mu\mathpunct{}\nonscript\mkern-\thinmuskip{;}
\mskip6muplus1mu\relax} 

\newcommand{\omegab}{\bar{\omega}}
\newcommand{\gammab}{\bar{\gamma}}
\newcommand{\betab}{\bar{\beta}} 
\newcommand{\ugamma}{\underline{\gamma}}

\newcommand{\ddd}{\mathbf{d}}

\newcommand{\fe}{\mathfrak{e}}
\newcommand{\fp}{\mathfrak{p}}
\newcommand{\fC}{\mathfrak{C}}

 
\tableofcontents

\vspace{20pt}

\large  

\section{Introduction}

It is known that the dynamics of Einstein's gravity can be represented in three different geometric ways using the Riemannian curvature, torsion tensor, 
or non-metricity tensor \cite{koivisto,capoz22}. It naturally leads to the question if the alternative formulations are better compatible with the conceptual framework of quantum theory and,  therefore, more suitable for the quantization of gravity, and whether the resulting quantum theories of gravity are equivalent or not. 

In this paper, we concentrate on the teleparallel formulation of gravity (see, e.g., \cite{per-rev, per-book, maluf-rev}) 
with a nonvanishing torsion tensor and vanishing curvature and nonmetricity. The underlying geometry is known as 
 the  Weitzenb\"ock geometry 
 (see \cite{sauer,cartan} for historical reviews)  
 and it replaces the Riemannian geometry of space-time in the standard formulation of general relativity. 
 
 The idea that teleparallel gravity can be a potentially favorable framework for the quantization of gravity and reconciliation of gravitation with quantum mechanics  has been put forward in 
 \cite{per-q3,per-q2,per-q1}. However, the potential advantages of teleparallel formulations,  
such as the  ability to define the gravitational energy-momentum tensor \cite{per-rev,per-book,maluf-energy} and the relation to the gauge theory of the translation group \cite{cho,hashir} 
 have not yet helped to address and resolve fundamental conceptual and technical problems of quantum gravity despite  numerous papers on the canonical Hamiltonian formulation 
\cite{maluf94,maluf99,capoz21,canontp1,blixt,canontp2} 
and  Ashtekar-like formulation \cite{warsaw,perim,mielke,okol} of the teleparallel gravity.   

In our opinion, all those problems, such as the mathematical definition of the Wheeler-De Witt equation 
 and its would-be teleparallel analog,  the problem of time, and the 
problem of the operational meaning of the ``wave function of the Universe" in quantum cosmology, or the problem of the  classical limit in loop quantum gravity, are rooted in the standard methods of canonical quantization based on the assumption of the global hyperbolicity of spacetime and an infinite dimensional extension of the formalism of quantum mechanics to field theory and, therefore, they will inevitably propagate to the corresponding approaches to quantization applied to the teleparallel theories of gravity.  
 
 The approach of precanonical quantization \cite{ik2,ik3,ik4,ik5,ik5e} based on the De Donder-Weyl (DW) Hamiltonian-like formulation in the calculus of variations \cite{kastrup,dw}, which does not require space-time decomposition,  and a generalization of the Poisson brackets to this formulation found in \cite{mybracket1,mybracket2,mybracket3,ik5}  (see also \cite{helein1,helein2,helein3,mybracket-pc,paufler1}), 
 were developed as a response to these problems. The generalized Poisson brackets are defined on horizontal differential forms on a finite-dimensional bundle over spacetime whose sections are field configurations. The standard Poisson brackets used for canonical quantization are defined on the functionals of the field configurations. The 
brackets of forms lead to the  Gerstenhaber algebra structure \cite{mybracket3,ik5} as a generalization of the Poisson algebra of the 
standard brackets of functionals.  
 
Precanonical quantization of fields is based on quantization of the Heisenberg-like subalgebra of the above-mentioned Gerstenhaber algebra 
according to the Dirac quantization rule  (\cite{ik3,ik4,ik5,ik5e}. It results in 
 a hypercomplex generalization of the formalism of quantum mechanics to field theory where the wave functions and operators are valued 
 in the  complexified Clifford algebra of spacetime. The latter naturally emerges from the precanonical quantization of Poisson-Gerstenhaber brackets of the dynamical variables represented by differential forms. 
The resulting quantum field theory is described in terms of the Clifford-algebra-valued precanonical wave functions on the finite-dimensional bundle of field variables $\phi^a$ over space-time $x^\mu$, viz., 
\beq 
\Psi (\phi^a,x^\mu)= \psi + \psi_\mu\gamma^\mu + \frac{1}{2!}\psi_{\mu_1\mu_2}\gamma^{\mu_1\mu_2}
+ \frac{1}{3!}\psi_{\mu_1\mu_2\mu_3}\gamma^{\mu_1\mu_2\mu_3}  + 
\frac{1}{4!}\psi_{\mu_1\mu_2\mu_3\mu_4}\gamma^{\mu_1\mu_2\mu_3\mu_4} , 
\eeq 
where $\gamma^\mu, \gamma^{\mu_1\mu_2}, \gamma^{\mu_1\mu_2\mu_3}, \gamma^{\mu_1\mu_2\mu_3\mu_4}$ are a 16-dimensional basis of the Dirac algebra 
and the components $\psi,\psi_\mu,...,\psi_{\mu_1\mu_2\mu_3\mu_4}$ all depend on $\phi^a$ and $x^\mu$. 
In the quantum theory of gauge fields $A_\mu^a$, the precanonical wave function is 
$\Psi (A^a_\mu, x^\mu)$ \cite{ik5e,ikym1,my-ymmg,ikym3},  and in the quantum theory of metric gravity, the precanonical wave function is $\Psi (g^{\mu\nu}, x^\mu)$ \cite{ikmetr1,ikmetr2,ikmetr3,ikmetr4}. Precanonical wave functions obey 
the analog of the Schr\"odinger equation 
\cite{ikmetr1,ikmetr2,ikmetr3,ikmetr4,ikan1,ikan2,ikan3,ikan4,ikan5}
\beq  \label{pseq}
i\hbar\varkappa \what{\slashed \nabla} \Psi = \what{H} \Psi ,
\eeq
where $\what{H}$ is the operator of the DW Hamiltonian function, which is the space-time symmetric analog of the canonical Hamiltonian in the DW Hamiltonian-like formulation of fields,  and 
$\what{\slashed \nabla}$ is the quantum version of the Dirac operator on the space-time. This equation can be viewed as a multidimensional generalization of the usual Schr\"odinger equation that treats all space-time variables on equal footing. 
The hat over the symbol of the Dirac operator means that this operator includes tetrads and spin connection coefficients which, 
in the context of quantum gravity, can be represented by differential operators.
  The parameter $\varkappa$ in (\ref{pseq}), which appears on purely dimensional grounds,  is an ultraviolet quantity of the dimension of the inverse $(n-1)-$dimensional  spatial volume (in $n$ space-time dimensions).  
In one-dimensional space-time  that corresponds to quantum mechanics this parameter is just a dimensionless multiplier that is absorbed in the definition of the time variable. 

Obviously, the precanonical description of quantum fields in terms of Clifford-algebra-valued
precanonical wave functions on a finite-dimensional space of field variables and space-time variables, $\Psi (\phi^a,x^\mu)$, 
is very different from the already conventional representations of quantum fields in terms of functionals of initial field configurations, 
as in the functional Schr\"odinger representation, or operator-valued distributions, as in the Heisenberg representation,  or nets of algebras of local observables, as in algebraic QFT. 
Nevertheless, it has been  shown that in the limiting case when 
$\varkappa$ goes over to the 
un-regularized 
total volume of the momentum space, that
is $\delta(\mathbf{0})$, the Dirac delta function at equal spatial points, 
the standard functional Schr\"odinger representation of QFT emerges from the precanonical Schr\"odinger equation (\ref{pseq})  \cite{ik-pla,iks1,iks2}. 
This relation between the precanonical formulation and the functional Schr\"odinger representation of standard QFT has been proven to hold  for scalar fields in Minkowski space-time \cite{iks1,iks2} and curved space-times \cite{iksc1,iksc2,iksc3}, and for quantum gauge fields 
 \cite{ikym3}. 
It can be understood as a consequence of the probabilistic interpretation of the precanonical wave function - as the probability amplitude of detecting the field value $\phi$ at the space-time point $x$ - and the probabilistic interpretation of the Schr\"odinger wave functional - as the probability amplitude of observing the field configuration $\phi(\bx)$ on the hypersurface of constant time $t$, which allows us  
 to understand the Schr\"odinger wave functional $\BPsi([\phi(\bx)], t)$  
 as a continuous product (or product integral) over all spatial points $\bx$ of precanonical wave functions restricted to the configuration 
$\phi = \phi(\bx)$, i.e. $\Psi (\phi = \phi(\bx), \bx, t)$. 
To be more precise, 
 the latter has to be inserted in the product integral after it has been transformed from the representation with the diagonal $\hat{H}$ to the representation with the diagonal energy density operator 
$\hat{T}{}^0_0 := \what{\der_0 \phi p^0_\phi - L} = \hat{H} - \der_i \phi\, \hat{p}_\phi^i$, where $p_\phi^\mu := \frac{\der L}{\der \der_\mu \phi}$ are the polymomenta and $H := \der_\mu \phi\, p^\mu_\phi - L$  is the definition of the DW Hamiltonian function from the Lagrangian $L = L (\phi, \der_\mu \phi, x^\nu)$.   Note that $\mu,\nu,... $ are spatiotemporal indices, and $i, j,... $ are spatial indices.  
  
  In terms of the new variables of the DW Hamiltonian formulation: $p^\mu_\phi$ and $H (\phi, p^\mu_\phi,x^\mu)$,  the Euler-Lagrange equations corresponding to $L$ take the form of the DW Hamiltonian field equations
\begin{align}  \label{dwheq}  
\begin{split}
 \bigg( \der_\mu \phi - \frac{\der H}{\der p_\phi^\mu} \bigg)  \bigg|_\sigma  &=0 , \\ 
 \left. \left( \der_\mu p_\phi^\mu + \frac{\der H}{\der \phi} \right)  \right|_\sigma &= 0 , 
 \end{split}
\end{align}
where ${ |}_\sigma$ denotes a restriction to the section $\sigma: \phi=\phi(x), p^\mu_\phi = p^\mu_\phi (x) $ of the bundle of the polymomentum phase space with the coordinates $(\phi, p^\mu_\phi)$ over the space-time with the coordinates $x^\mu$  (see, e.g.,  \cite{ik5} for a more intrinsic definition).  
The symbol $\phi$ here may also mean a multicomponent object together with all its internal and space-time indices, such as  
$\phi^a$ or $A_\mu$.
 When the regularity condition  $\det \big( \frac{\der^2 L}{\der \phi_\mu \der \phi_\nu} \big) \neq 0$ is not satisfied, the DW formulation is singular  
 or constrained in the sense of DW theory 
 and one needs a proper treatment of constraints within the DW Hamiltonian theory. The approach based on the notion of the polysymplectic structure and a Dirac-like generalization of Poisson-Gersten\-haber brackets of differential forms representing both dynamical variables and  constraints has been developed in \cite{mydirac}. Further development and applications of the approach can be found in \cite{ikan1,ikan2,mx1,mx2,mx3,mohj1,mohj2}.

 Note also that the consideration of constrained DW Hamiltonian 
systems from the perspective of the jet bundle geometry of classical fields can be found in several recent papers such as \cite{rr2023,marco22}. These authors do not use the classical Dirac's procedure based on Poisson brackets. Nevertheless,  their analysis of the geometry of the constrained DW Hamiltonian 
systems may help 
 to understand the geometric foundations of our approach and the reduction of the polymomentum phase space it naturally leads to in the case of vielbein general relativity in our previous 
  papers \cite{ikan1,ikan2,ikan3,ikan4,ikan5}  and its teleparallel equivalent in this 
  paper (see also \cite{mx1,mx2,mx3,mohj1,mohj2}).   

Note also that the DW Hamiltonian formulation of field dynamics in (\ref{dwheq}) leads to a space-time symmetric  analog of the Hamilton-Jacobi theory dating back to the work of Volterra, De Donder and Weyl \cite{dw,kastrup}. In the context of gravity, the latter has been studied in \cite{mohj1,horava,riahi}. The relation between the DW Hamilton-Jacobi equation and the classical limit of the precanonical Schr\"odinger equation for scalar fields has been discussed in \cite{ik3,guiding}. 
 
In this paper, we discuss the precanonical quantization of the teleparallel equivalent of tetrad general relativity. We proceed as follows.  In Section 2, we present  the Palatini formulation of tetrad TEGR and its DW Hamiltonian formulation, which is a constrained theory. We analyze the constraints using the method developed by us for singular DW Hamiltonian systems and calculate the generalized Dirac brackets of the fundamental variables. This leads to a reduction of the polymomentum phase space of the Palatini formulation to the space of tetrads and their 
polymomenta, and the 
expressions 
of the DW Hamiltonian function and the auxiliary fields of the Palatini formulation in terms of the variables of the reduced polymomentum phase space. Then, in Section 3, we quantize the resulting DW Hamiltonian system on the reduced polymomentum phase space according to the  modified Dirac quantization rule, find the representations of operators up to an operator ordering ambiguity, and present the scalar product with the invariant integration measure over tetrad coefficients. 
The operator ordering 
of the DW Hamiltonian operator
that is consistent with the weight factor in the integration measure in the scalar   
product, 
the related emergence of the cosmological constant, and the estimation of its numerical value, are discussed in 
Section 4. Section 5 presents conclusions. 


\section{The DW Hamiltonian formulation of TEGR in tetrad variables} 

\subsection{The Palatini formulation of TEGR in tetrad variables} 

TEGR is a teleparallel equivalent of general relativity which is formulated in terms of the  Weitzenb\"ock geometry with nonvanishing torsion instead of the Riemannian geometry of general relativity with a nonvanishing curvature and vanishing torsion. 
The field equations of TEGR are the same as Einstein's field equations of general relativity despite a different underlying geometry. 

In our previous papers \cite{ikan1,ikan2,ikan3,ikan4,ikan5},  a procedure of precanonical quantization of vielbein general relativity in the Palatini formulation has been presented. Here we will show how this approach can be extended to TEGR. It is natural to use an appropriate Palatini tetrad formulation of TEGR for the purpose of comparison of two quantum theories.  Among different formulations of TEGR that have been considered in numerous papers (see, e.g., \cite{koivisto-pal, maluf2}) 
we found the Palatini formulation  by Maluf \cite{maluf} to be the most suitable for our goal.

In Maluf's formulation, the tetrad components $e^\mu_a$ and the auxiliary variables $f_{abc}= - f_{acb}$ 
 with  antisymmetric last two indices are the independent field variables, and the  Lagrangian density has the form 
\beq  \label{lagrtp}
\mathfrak{L} = \frac{1}{16\pi G} \fe \Phi (f)^{abc} \left(f_{abc}-2T_{abc}\right) , 
\eeq
where 
\begin{align} \label{fff}
\begin{split}
\Phi (f)_{abc}&:=\frac14 ( f_{bac} + f_{abc} -f_{cab}) + \eta_{a[c} f_{b]} ,   \\ 
f_c&:= \eta^{ab}f_{abc},   
\end{split}
\end{align} 
and 
\begin{align}\label{tors}
\begin{split}
T^{c}{}_{ \mu\nu} &:= \der_\mu e^c_\nu - \der_\nu e^c_\mu , 
\\
T^c{}_{ab} &:= e^\mu_a e^\nu_b T^c{}_{\mu\nu} , 
\quad 
T_c := T^a{}_{ac}.  
\end{split}
\end{align}
Here and in what follows, the Gothic letters like 
 $\fe := \det (e^a_\mu)$ denote densities. 

By restricting the Lagrangian density to a field configuration 
\beq 
e^c_\nu =  e^c_\nu (x), \quad 
\der_\mu e^c_\nu = \der_\mu e^c_\nu (x), \quad 
f_{abc} = f_{abc} (x) 
\eeq 
and integrating over $x^\mu$-s, $\mu = 0,1,2,3,$ we obtain the action functional. 

The variation with respect to the auxiliary fields $f_{abc} (x)$ leads to the equation 
\beq  \label{deltaf}
\Phi (f)^{abc} = \Phi (T)^{abc} , 
\eeq 
which is equivalent to \cite{maluf94}
\beq  \label{feqt}
f_{abc} = T_{abc}  , 
\eeq
i.e. the off-shell auxiliary fields $f_{abc}$ are identified  with the torsion tensor (\ref{tors}) on-shell. 

The variation  with respect to 
tetrad fields $e^c_\mu (x)$ 
and the use of (\ref{feqt}) 
leads to the field equations 
\beq
e_{a\sigma}e_{b\mu}\der_\tau \left(\fe \Phi^{b\sigma\tau}(T)\right) 
- \fe  \big( \Phi^{b\sigma}{}_a(T)  T_{b\sigma \mu} 
- \frac14 e_{a\mu} T_{cbd} \Phi^{cbd}(T)\big) =0  
\label{deltae}
\eeq 
and reproduces the tetrad form of the vacuum Einstein equations \cite{maluf94}. 

\subsection{The DW Hamiltonian formulation and the constraints}

The polymomenta densities derived from the Lagrangian density (\ref{lagrtp}) have the form 
\begin{align}
\mathfrak{p}{}^\mu_{f_{abc}}&:= \frac{\der \mathfrak{L}}{\der \der_\mu f_{abc}}=0, 
 \\
\mathfrak{p}{}^\mu_{e^a_\nu} &:= \frac{\der \mathfrak{L}}{\der \der_\mu e^a_\nu}
  =  - \frac{1}{4\pi G}\fe \Phi(f)_a{}^{bc} e^{[\mu}_b e^{\nu]}_c .
\end{align}
The expressions show that the Legendre transformation from the first derivatives of our field variables $f$ and $e$ to their corresponding 
polymomenta is singular and we need an analysis of constraints adapted to the DW Hamiltonian formulation. 
In spite of many papers trying to treat singular DW Hamiltonian systems from the point of view of the geometry of jet bundles 
(see, e.g., \cite{deleon-book}), the approach suitable for quantization, which is similar to Dirac's treatment of constraints, was 
developed in \cite{mydirac} and then applied in a refined form to the analysis of the singular DW Hamiltonian system derived from the Palatini formulation of Einstein's vielbein gravity \cite{ikan1,ikan2,ikan3,ikan4,ikan5} (see also \cite{mx1,mx2,mx3,mohj1,mohj2} for further applications). 

Using the language and methods of \cite{mydirac}, we identify the primary constraints (in the sense of DW Hamiltonian theory): 
\begin{align}
\mathfrak{C}{}^\nu_{f_{abc}} & :=\mathfrak{p}{}^\nu_{f_{abc}} \approx 0, \label{cf}\\
 \mathfrak{C}{}^\mu_{e^a_\nu}\; & :=  \mathfrak{p}{}^\mu_{e^a_\nu} +  \frac{1}{4\pi G}\fe \Phi(f)_a{}^{bc} e^{[\mu}_b e^{\nu]}_c \approx 0 , 
\label{ce}
\end{align}
and the primary DW Hamiltonian density: 
\beq \label{dwhprim}
{\fe H} := 
\mathfrak{p}^\mu_{f_{abc}} \der_\mu f_{abc} + \mathfrak{p}^\mu_{e^a_\nu} \der_\mu {e^a_\nu} 
- \mathfrak{L} 
\approx  - \frac{1}{16\pi G} \fe \Phi (f)^{abc} f_{abc} ,  
\eeq
where the weak equality takes into account the constraints (\ref{cf}). 
On the surface of 
DW constraints (\ref{ce}) the latter can be written in the form 
\beq \label{dwhprimc}
{\fe H} :
\approx \frac14 \mathfrak{p}^\mu_{e^a_\nu} e_\mu^b e_\nu^c f^a{}_{bc}  . 
\eeq


\subsection{The (pre-)polysymplectic formulation and the Poisson-Gerstenhaber brackets}

Now, in order to analyze the brackets of constraints and their preservation, let us recall, based on a general theory of the polysymplectic structure and the Poisson-Gerstenhaber brackets in the DW Hamiltonian formulation \cite{mybracket3},  
that the unconstrained extended  polymomentum phase space 
of variables 
$(f_{abc},  e_\mu^a, \mathfrak{p}^\mu_{f_{abc}},  \mathfrak{p}^\mu_{e_\mu^a},x^\mu)$ 
or, in our shorthand notation, 
$(f, e, \mathfrak{p}^\mu_{f}, \mathfrak{p}^\mu_e, x^\mu)$   
is naturally~equip\-ped with the \mbox{(pre-)}polysymplectic structure given by 
a representative of the equivalence class of forms modulo semi-basic $4-$forms \cite{ik5}  
\beq \label{preps}
\Omega = d \mathfrak{p}{}^\mu_f \we d f \we \upsilon_\mu +
d\mathfrak{p}{}^\mu_e \we d e \we \upsilon_\mu ,  
\eeq
 where $\upsilon_\mu$ denotes the basis of $3-$forms in $4$-dimensional space-time: 
\beq 
\upsilon_\mu := \der_\mu \inn \left (dx^0\we dx^1\we dx^2 \we dx^3 \right). 
\eeq  
$\Omega$ maps (a certain class of) semi-basic 
$p-$forms, $p=0,1,2,3$,   
\beq 
\ff{p} := \frac{1}{p!} F_{\nu_1 ... \nu_p} (f, e, \mathfrak{p}^\mu_{f}, \mathfrak{p}^\mu_e, x^\mu) 
\, dx^{\nu_1}\we ...\we dx^{\nu_p} ,
\eeq
to (an equivalence class of) $(4-p)-$multivector fields $X_{\mbox{\scriptsize $\ff{p}$}}$  (modulo the multivectors $X_0$ from the kernel: \mbox{$X_0 \inn\ \Omega =0$}), i.e.  
\beq  \label{map}
X_{\mbox{\scriptsize $\ff{p}$}} \inn\ \Omega = d \ff{p} .  
\eeq
This allows us to  define a bracket operation on forms for which the map (\ref{map})  exists \cite{mybracket1,mybracket2,mybracket3,ik5} 
\beq \label{gbr}
\pbr{\ff{p}_1}{\ff{q}_2} := (-1)^{p} X_{\mbox{\scriptsize $\ff{p}$}} \inn\ d\ff{q}_2  . 
\eeq
The product operation on those forms is given by the co-exterior product operation $\bullet$: 
\beq \label{bullet}
F\bullet G := *^{-1}(*F\we*G) .  
\eeq
The bracket (\ref{gbr}) and the product operation (\ref{bullet}) lead to the Poisson-Gerstenhaber structure on semi-basic 
forms. 

In fact, a more general class of ``Poisson forms" and their brackets were defined in the literature \cite{paufler1}. However, the  bracket operation on these forms is graded Lie, not Poisson, as there is no product operation on the space of ``Poisson forms." 

Using the definitions above we can now calculate the brackets of $3-$forms constructed from the constraints (\ref{cf}), (\ref{ce}): 
\beq 
\mathfrak{C}{} _f:= \mathfrak{C}{}^\mu_{f}\upsilon_\mu  ,\quad 
\mathfrak{C}{}_e:=  \mathfrak{C}{}^\mu_{e}\upsilon_\mu .    
\eeq
A simple calculation yields 
\beqa
&\pbr{\mathfrak{C}{}_f}{\mathfrak{C}{}_{f'}} &\!\!\! =: \mathfrak{C}{}_{f f'} = 0 , \\
&\pbr{\mathfrak{C}{}_{e^a_\mu}}{\mathfrak{C}{}_{e^d_\nu}} &\! \!\! = \frac{1}{4\pi G} \der_{e^a_\mu}\left( \fe \Phi(f)_d{}^{bc} e^{[\alpha}_b e^{\nu]}_c \right)
\upsilon_\alpha =: \mathfrak{C}{}_{e^a_\mu e^d_{\nu}}
 , 
\label{cece} \\   
&\pbr{\mathfrak{C}{}_{e^d_\nu}}{\mathfrak{C}{}_{f_{abc}}} \ &\!\!\! = -\frac{1}{4\pi G}\fe e^{[\mu}_g e^{\nu]}_h 
\frac{\der\Phi(f)_d{}^{gh}}{\der {f_{abc}}} \upsilon_\mu
=: \mathfrak{C}{}_{e^d_\nu f_{abc}}
, 
\eeqa
where $ \mathfrak{C}{}_{e e'} $ depend on $f$ and $e$ and  $ \mathfrak{C}{}_{e f} $ depend only on $e$. 
Thus the block structure of the matrix of brackets of constraints 
$\mathfrak{C}_{UV}^\mu\upsilon_\mu$, 
 where the $(4\times 6 +4\times 4)=40-$dimensional
 indices $U,V$  enumerate the constraints (\ref{cf}), (\ref{ce}) and run over all the components of $f$ and $e$, 
 has the form 
\beq  \mathfrak{C}_{UV} = 
\begin{bmatrix}   
    0      & \; \mathfrak{C}_{f e } \\
    \mathfrak{C}_{e f} 
    & \; \mathfrak{C}_{e e'} 
\end{bmatrix} ,   
\eeq
where 
the rectangular submatrices $\mathfrak{C}_{e f } = -  \mathfrak{C}_{f e}^T$ and the matrix elements $\mathfrak{C}_{UV}$ are densities of weight $+1$.

\subsection{The generalized Dirac brackets and the reduced polymomentum phase space}

As $\mathfrak{C}_{e f}$ is not weakly vanishing, the  constrained DW Hamiltonian system obtained from the Palatini Lagrangian (\ref{lagrtp}) 
is similar to the second-class constrained Hamiltonian systems. A proper generalization of\ Dirac brackets to second-class constrained DW Hamiltonian systems is defined on forms $F$ and $G$ of degree $3$ and $0$ \cite{mydirac,ikan1}: 
\beq \label{pdir}
\pbr{F}{G}{}^D := \pbr{F}{G}{} - \pbr{F}{\mathfrak{C}_U}{}\bullet \big ( \mathfrak{C}_{UV}^{\sim 1} \we \pbr{\mathfrak{C}_V}{G}{} \big), 
\eeq
where $\mathfrak{C}^{\sim 1}_{UV} =\mathfrak{C}^{\sim 1}_{\mu UV} dx^\mu $ is the generalized pseudoinverse matrix 
whose matrix elements are one-forms  such that (cf. \cite{penrose,pseudoinv})
\beq
\mathfrak{C}_{UU'}\bullet \mathfrak{C}^{\sim 1}_{U'V'} \!\we \mathfrak{C}_{V'V} = \mathfrak{C}_{UV},  
\eeq 
and the summation over the repeated indices $U'$ and $V'$ is assumed. 

Note that the appearance of the pseudoinverse in our definition of generalized Dirac brackets is justified by the nonvanishing multivector kernel of the polysymplectic forms like  (\ref{preps}). It is, however,  less justified in the canonical formalism where the vector kernel of symplectic forms is vanishing, when the pseudoinverse is used as an attempted shortcut replacing the standard analysis of constraints {\em \`a la} Dirac (c.f. \cite{canontp2}).

The matrix $\mathfrak{C}^{\sim 1}_{UV}$ whose components are densities of weight $-1$ 
 has the block structure 
\beq\label{pseinv}
\mathfrak{C}^{\sim 1}_{UV} = 
\begin{bmatrix}  
    \mathfrak{C}^{\sim 1}_{f f'} \!    & \; \mathfrak{C}^{\sim 1}_{f e}   \\
    \mathfrak{C}_{ ef }^{\sim 1}     & \; 0  \!
\end{bmatrix}, 
\eeq
where  $\mathfrak{C}_{e f }^{\sim 1}= - \mathfrak{C}^{\sim 1}_{f e}{}^T$ and 
\begin{align}
\mathfrak{C}^{\sim 1}_{\mu e f} \mathfrak{C}^{\mu}_{f e'} &= \delta_{ee'}, 
\nn \\ 
\mathfrak{C}^{\sim 1}_{\mu fe } \mathfrak{C}^{\mu}_{ef'} & = \delta_{ff'}, \label{cinver}
 \\ 
\mathfrak{C}^{\sim 1}_{\mu f e'} \mathfrak{C}^{\mu}_{e'e}  + \mathfrak{C}^{\sim 1}_{\mu f f'} \mathfrak{C}^{\mu}_{f'e} & =0  .
\nn 
\end{align}


\subsubsection{The Dirac brackets of fundamental variables}

Now we are ready to calculate the generalized Dirac brackets of fundamental variables, i.e. the forms constructed from 
 the auxiliary variables $f$, the tetrad coefficients $e$, and their respective polymomenta. Using the notation $\mathfrak{p}{}_e:= \mathfrak{p}{}^\mu_e \upsilon_\mu$ for the 3-forms of the polymomenta of tetrads we obtain 
\begin{align}
\pbr{\mathfrak{p}{}_e}{e'}{}^D &= 
 \pbr{\mathfrak{p}{}_e}{e'}{}=\delta_{ee'}  , 
\label{pee}
\\
\pbr{\mathfrak{p}{}^\mu_e}{e'\upsilon_\nu}{}^D 
& =  \pbr{\mathfrak{p}{}^\mu_e}{e'\upsilon_\nu}{}=\delta_{ee'}\delta^\mu_\nu  , 
\label{pmuee}
\\
\pbr{\mathfrak{p}{}_e}{e'\upsilon_\mu}{}^D &= 
 \pbr{\mathfrak{p}{}_e}{e' \upsilon_\mu}{} = \delta_{ee'} \upsilon_\mu ,\\
\pbr{\mathfrak{p}{}_e}{\mathfrak{p}{}_{e'}}{}^D &=0 ,\\
\pbr{e \upsilon_\mu}{e'_{}}^D &
=  \mathfrak{C}^{\sim 1}_{\mu ee'} = 0 . \label{ee}
\end{align} 
Those brackets coincide with the brackets we have obtained earlier for scalar fields \cite{mybracket1,mybracket2,mybracket3}, and they can be easily quantized  in terms of Clifford-algebra-valued operators \cite{ik3,ik4,ik5e}. 
 
A similar calculation for field variables $f$ and their polymomenta $\mathfrak{p}{}_f:= \mathfrak{p}{}^\mu_f\upsilon_\mu$  
yields 
\begin{align}\label{pff}
\begin{split}
\pbr{\mathfrak{p}{}_f}{f'}{}^D &= \pbr{\mathfrak{p}{}_f}{f'}{} - \pbr{\mathfrak{p}{}_f}{ \mathfrak{C}_e}\bullet (  \mathfrak{C}^{\sim 1}_{e f''} 
\we \pbr{\mathfrak{C}_{{f''}}}{f'})
 \\
& = 
\delta_{ff'} - \mathfrak{C}_{fe}\bullet \mathfrak{C}^{\sim 1}_{e f''} \delta_{f'' f'} 
= \delta_{ff'} - \mathfrak{C}^\mu_{fe} \mathfrak{C}^{\sim 1}_{\mu e f'}
= 0,  \!\!\mbox{\!\!} 
\\ 
\pbr{\mathfrak{p}{}_f}{f'\upsilon_\mu}{}^D  &= 0 , \\
 \pbr{\mathfrak{p}{}^\mu_f}{f'\upsilon_\nu}{}^D &=0, \\
 \pbr{\mathfrak{p}{}_f}{ \mathfrak{p}{}_{f'}}{}^D &=0 . 
 \end{split}
\end{align}
At the same time, 
\begin{align}
 \pbr{f \upsilon_\mu}{f'_{}}^D &
 = \mathfrak{C}^{\sim 1}_{\mu ff'} , 
 \label{ff} \\
 \pbr{\mathfrak{p}{}_e}{f}{}^D &=
\pbr{\mathfrak{p}{}_e}{f}{} -  \pbr{\mathfrak{p}{}_e}{\mathfrak{C}_{e'}} \bullet ( \mathfrak{C}^{\sim 1}_{e' f'} \we \pbr{\mathfrak{C}_{f'}}{f} ) 
\nn \\ & 
= - \frac{1}{4\pi G} \der_e \left(\fe{} \Phi(f)_b{}^{ac} e{}_a^\alpha e_c^\nu \right)  \mathfrak{C}^{\sim 1}_{\alpha e^b_\nu f} = 
- \mathfrak{C}^\alpha_{e e'} \mathfrak{C}^{\sim 1}_{\alpha e' f}  , 
\label{pef}\\
\pbr{e \upsilon_\mu}{f_{}}^D &
= \mathfrak{C}^{\sim 1}_{\mu ef} , 
\label{ef}  \\ 
 \pbr{\mathfrak{p}{}_f}{\mathfrak{p}{}_{f'}}{}^D & = 
\pbr{\mathfrak{p}{}_f}{e}{}^D  
= 
\pbr{\mathfrak{p}{}_f}{\mathfrak{p}{}_{e}}{}^D = 0 . 
\label{pfpe} 
 \end{align}

\subsubsection{The polysymplectic reduction}

The above Dirac brackets indicate that the original polymomentum phase space of variables 
$(f, \mathfrak{p}_f, e, \mathfrak{p}_e)$ is reduced 
to the space of tetrad coefficients $e^a_\mu$ and their polymomenta $\mathfrak{p}{}^\mu_{e^\nu_a}$, and the auxiliary field variables 
 $f_{abc}$, whose mutual brackets are not vanishing ({\ref{ff}}),
 are becoming functions on the reduced polymomentum phase space:   $f = f(e ,\mathfrak{p}_e )$. 
The reduced polymomentum phase space is 
equipped with the reduced polysymplectic structure given by 
\beq \label{omegar} 
\Omega_\mathrm{red} = d\mathfrak{p}{}^\mu_{e^a_\nu} \we d e^a_\nu \we \upsilon_\mu , 
\eeq  
which reproduces the brackets (\ref{pee})--(\ref{ee}) as the standard Poisson-Gerstenhaber brackets. 

Let us note that our notion of the polysymplectic reduction based on a generalization of Dirac brackets is different from other constructions using the same term but a different understanding of the polysymplectic structure and the related brackets (c.f. \cite{mcclain,rr-psred,blacker}). Both their work and our first paper on the subject \cite{mybracket1} are inspired by G\"unther's work in the 1980-es \cite{guenther}. A detailed review of different relevant geometric structures in classical field theory can be found in the book \cite{deleon-book}.  In spite of all these developments, it is our understanding of the polysymplectic structure as  the equivalence of class of forms modulo the coordinate-dependent horizontal contribution of the top degree \cite{mybracket2,ik5} that allows defining a proper analog of the Poisson brackets in the DW Hamiltonian theory, which does not depend on the choice of the representative of the equivalence class of polysymplectic forms, and the Dirac brackets in the constrained  DW Hamiltonian systems with second class constraints (in the sense of the DW Hamiltonian theory). This is the only framework so far which has allowed us to formulate of the foundation of precanonical quantization in field theory based on the DW Hamiltonian formulation.  A more naive definition of the Poisson brackets in the DW Hamiltonian theory (c.f. \cite{my-dkp} and the references therein) which different authors tried to use recently 
(c.f. \cite{struck,mcclain,blacker,fernandez}) 
 do not lead to a  natural representation of the DW Hamiltonian field equation in terms of the bracket and to proper 
Leibniz and Jacobi properties which allow 
us 
to use the Dirac's quantization rule 
by replacing the classical brackets with quantum commutators.

\subsection{\texorpdfstring{\mbox{$f_{abc}$} and \mbox{$H$} on the reduced polymomentum phase space}{fabc and H on the reduced polymomentum phase space}}

As we have noticed, the bracket (\ref{ff}) indicates that the auxiliary fields $f_{abc}$ are functions of $e$ and $\mathfrak{p}{}_e$.
To obtain the explicit expression of $f_{abc} (e,\mathfrak{p}{}_e)$ let us start from the consequence of equation (\ref{fff})
\begin{align}
\begin{split}
f_{abc} + \eta_{ac} f_b -\eta_{ab} f_c  &= 2 \Phi_{bac} - 2 \Phi_{cab} , \nn \\
f_c &= 
-  \Phi_c , \nn 
\end{split}
\end{align}
whence it follows 
\beq
\frac12 f_{abc} =  \Phi_{bac} -  \Phi_{cab} 
- \frac12 (\eta_{ab}\Phi_c - \eta_{ac}\Phi_b) . 
\eeq
From (\ref{ce})  it follows 
\begin{align}
\begin{split}
 \Phi_{a}{}^{bc} &\approx {-4\pi G}\, e_{ [\mu }^b e^c_{ \nu  ]} {p}{}^\mu_{e^a_\nu} ,  \\
  \Phi^c &= \delta^a_b \Phi_{a}{}^{bc} \approx {-4\pi G}\, e^a_{[\mu} e^c_{\nu ]} p^\mu_{e^a_\nu} ,  
\end{split}
\end{align} 
where  the notation ${p}{}_e^\mu := \fe{}^{-1} \mathfrak{p}^\mu_e$ is introduced. 
Hence, 
\begin{align}  \label{fabc}
\begin{split}
 f_{abc} &\approx {8\pi G} \Big(   e_{a [\mu } e_{b| \nu  ]} p^\mu_{e^c_\nu} -   e_{a [\mu } e_{c| \nu  ]} p^\mu_{e^b_\nu}
  \\  & 
  + 
 \frac12 \left( \eta_{a b} e^d_{[\mu} e_{c|\nu ]} p^\mu_{e^d_{\nu}} 
  - \eta_{a c} e^d_{[\mu} e_{b |\nu ]} p^\mu_{e^d_{\nu}} \right) \Big)  
  \end{split}
\end{align}
or, in a shorter form, 
\beq \label{fabc2}
f_{abc} \approx 
 {16\pi G} \Big(   
 e_{a \mu } e_{[b \nu } {p}{}^{[\mu}_{e^{c]}_{\nu]}}  
+ 
\frac12 \eta_{a [b} e^d_{\mu} e_{c] \nu } p^{[\mu}_{e^d_{\nu]}} 
   \Big)  . 
\eeq

Now we can write  the DW Hamiltonian density $\fe H$  (\ref{dwhprimc}) as a function on the reduced polymomentum phase space, viz.,  
\begin{align} \label{hdwr}
\begin{split}
 H  &\approx \frac14 p^\mu_{e^a_\nu} e_\mu^b e_\nu^c f^a{}_{bc} 
\approx 
 {2\pi G}  p^\alpha_{e^a_\beta} e_\alpha^b e_\beta^c \left( e_{ [\mu }^a e_{b |\nu  ]} p^\mu_{e^c_\nu}  
  -   e_{ [\mu }^a e_{c|\nu]} 
 p^\mu_{e^b_\nu}  
+  
   \eta^a_{b} e^d_{[\mu} e_{c |\nu ]} p^\mu_{e^d_\nu}\right)
 \\
& 
\approx {4\pi G}  p^\alpha_{e^a_\beta} e_{[\alpha}^b e_{\beta]}^c \left(
  e_{ [\mu }^a e_{b |\nu  ]} p^\mu_{e^c_\nu}  
+  
\frac{1}{2}  \eta^a_{b} e^d_{[\mu} e_{c |\nu ]} p^\mu_{e^d_\nu} 
\right) . 
\end{split}
\end{align}

Introducing the notation $p^a_{e^b_\nu} := e_\mu^a p^\mu_{e^b_\nu},  \;
p_{\mu e^b_\nu} = g_{\mu\alpha}p^\alpha_{e^b_\nu}  $  we can also write 
\beq  \label{hdwr1}
\begin{aligned}
H  &\approx {\pi G}\Big( 2 e^c_\beta p_{\nu e^a_\beta} p^a_{e^c_\nu } 
-g_{\beta\nu} p^c_{e^a_\beta}p^a_{e^c_\nu}
-g_{\alpha\mu}e^c_\beta e^a_\nu p^\alpha_{e^a_\beta}p^\mu_{e^c_\nu} \\ 
& +  
\frac{1}{2}
\big( 
g_{\beta\nu}p^a_{e^a_\beta}p^b_{e^b_\nu} - 2 e^a_\beta p_{\nu e^a_\beta} p^b_{e^b_\nu}
  + g_{\alpha\mu}e^b_\nu e^a_\beta p^\alpha_{e^a_\beta}p^\mu_{e^b_\nu} 
\big) \Big)  .
\end{aligned} 
\eeq



\section{Precanonical quantization}

The fundamental brackets (\ref{pff})-(\ref{ff}) are quantized according to the modified Dirac's quantization rule
\beq
[\hat{A},\hat{B}] = -i\hbar \what{\mathfrak{e} \pbr{A}{B}^{ D}} .
\eeq 
This rule ensures that quantum representations of classical densities are density-valued operators. 
From (\ref{pee})  we  obtain
\beq  \label{pmuomop}
\what{\mathfrak{p}}{}_e := 
\what{\mathfrak{p}{}_e^\mu \upsilon_\mu} = -i \hbar\fe \der_e .
\eeq 

Then, from (\ref{pmuee})  it follows 
\beq \label{pmueop}
\what{\mathfrak{p}}{}_e^\mu  = -i\hbar\ka \fe \gamma^\mu \der_e   
\eeq
and 
\beq \label{omegaop}
\what{\upsilon}_\mu = \frac{1}{\varkappa }\gamma_\mu .  
\eeq
This representation assumes that the composition of Clifford-algebra-valued operators implies the symmetrized Clifford product (c.f.  \cite{ik3,ik5e}). 
The ultraviolet parameter $\varkappa$ in (\ref{omegaop}) has the physical dimension of the inverse 3-volume,  and it appears on dimensional grounds. 
The curved space-time Dirac matrices $\gamma^\mu$ are defined using the flat spacetime Dirac matrices 
$\ugamma^a$: 
\beq 
\gamma^\mu := e^\mu_a \ugamma^a  , 
\eeq 
\beq
\ugamma^a\ugamma^b+  \ugamma^b\ugamma^a = 2 \eta^{ab},  
\eeq 
 where $\eta^{ab}$ is the Minkowski metric. 

Obviously,  the operators in (\ref{pmuomop}), (\ref{pmueop}) are defined only up to an ordering of multiplicative operators $e$ and 
differential operators $\der_e$.  Their precise definition will depend on the scalar product 
to be defined below.

\subsection{\texorpdfstring{The operators of \mbox{$f_{abc}$} and \mbox{$H$}}{The operators of fabc and H}}

Now, using the representation of polymomenta in (\ref{pmueop}) and the classical expression of $f_{abc}$ (\ref{fabc})  
and $H$  (\ref{hdwr}), we obtain  
\begin{align}\label{fabcop}
\begin{split}
\hat{f}_{abc} & = - {8 \pi i  G}  \hbar\ka \Big( \big( \ugamma{}_a e_{[b|\nu} 
- e_{a\nu}\ugamma{}_{[b}\big)  \der_{e^{c]}_\nu} 
+ 
 \frac12 
 \big(\ugamma^d \eta_{a[b}e_{c]\nu} 
- \eta_{a[b}\ugamma{}_{c]} e^d_\nu   \big) \der_{e^d_\nu} \Big)  . 
\end{split}
\end{align}
Using the identity 
\beq
\gamma^\mu\ugamma{}_a +\ugamma{}_a \gamma^\mu = 2 e^\mu_a   
\eeq
and the observation that the operator of $g_{\mu\nu} p^\mu_e p^\nu_{e'}$  takes the form 
$-\hbar^2\ka^2\der_e\der_{e'}$,   similarly to the operator of 
$g_{\mu \nu} p^\mu_\phi p^\nu_{\phi}$  in scalar field theory given by $-\hbar^2\ka^2\der_{\phi\phi}$ 
(c.f. \cite{ik3,ik4,ik5e}),  we also obtain 
\begin{align} \label{hop}
 \what{H} = -\pi G \hbar^2\ka^2  \Big( e^c_\beta e^a_\nu - 
 \frac32 
 g_{\beta\nu} \eta^{ac}
- 
   \frac12 
   e^a_\beta e^c_\nu \Big)
 \der_{e^a_\beta}\der_{e^c_\nu}  .
\end{align}


\subsection{The precanonical Schr\"odinger equation for TEGR} 

It has been argued in our previous work (see, e.g.,  \cite{iksc1,iksc2,iksc3,ikmetr1,ikmetr2,ikmetr3,ikmetr4,ikan1,ikan2,ikan3,ikan4,ikan5}) 
that the precanonical Schr\"odinger equation for quantum gravity 
has the form  
\beq 
i \hbar\ka \what{\slashed\nabla}  \Psi \!=\! \what{H} \hspace*{-0.0em} \Psi  . 
\eeq 
In the context of quantum TEGR, 
$\what{\slashed\nabla}$ is   the quantized Dirac operator  in which 
 the spin connection is an operator acting via the commutator Clifford product on the 
 Clifford-algebra-valued wave function $\Psi(e,x)$, 
 \beq
 \what{\slashed\nabla} := 
 \gamma^\mu\left(\der_\mu+   
  \what{\omega}_{\mu}\mbox{$\stackrel{\leftrightarrow}{\vee}$}
 \right) , 
 \eeq 
where  $\gamma^\mu =  e^\mu_a \ugamma^a $
 and 
$ \what{\omega}_{\mu} =  \frac14 \what{\omega}_{\mu bc}{\ugamma}^{bc}$ 
is the operator of spin connection 
acting on the Clifford-valued precanonical wave function $\Psi$  by the commutator 
(antisymmetric) Clifford product $\stackrel{{\leftrightarrow}}{\vee}$  
\beq
{\ugamma}^{bc} \stackrel{\leftrightarrow}{\vee} \Psi := \frac12\left( {\ugamma}^{bc} \vee \Psi - \Psi \vee {\ugamma}^{bc} \right) = \frac12 \left[\ \ugamma^{bc}, \Psi\ \right] , 
\eeq
where $\vee$ is the Clifford algebra product  which, in the matrix representation of Clifford algebras,  is just the matrix product.  

  The equation for the  conjugate wave function  
  \beq
  \Psib := \ugamma{}^0\Psi^\dagger\ugamma{}^0  , 
  \eeq 
where ${}^\dagger$ denotes the standard 
conjugate transpose of the matrix, 
has the form  
\beq \label{pse-conj}
\ri\hbar\ka \Psib 
{\left( \stackrel{\leftarrow}{\der_\mu}  
+ \mbox{ $\stackrel{\leftrightarrow}{\vee}$} \what{\omegab}{}_\mu\right) } \gamma^\mu 
= - \hat{H} \Psib , 
\eeq
where 
$\what{\omegab}_\mu \!:=\!\ugamma{}^0\what{\omega}_\mu^\dagger\ugamma{}^0$,  
and a generalized Hermicity (or pseudo-Hermicity \cite{mostafa})  of $\hat{H}$ 
is assumed: 
\beq \label{gherm}
\hat{H} = \ {\overline{\!\!\hat{H}\!}},  
\quad {\overline{\!\!\hat{H}\!}} := \ugamma{}^0\hat{H}{}^\dagger\ugamma{}^0 
,    
\eeq
  where $\hat{H}{}^\dagger$ is the Hermitian conjugate of $\hat{H}$.  
   

 \subsubsection{The spin connection operator} 
 
 We still need to construct the operator of spin connection in precanonical Schr\"odinger equation. Classically, 
  the teleparallel spin connection is related to the contorsion \cite{pereira01,maluf03} 
 \beq
 \omega_{\mu ab} = - K_{\mu ab} = \half e^c_\mu\left(T_{a bc} + T_{ac b} - T_{c ab}\right) .
 \eeq
 For quantum representation, we need an  expression in terms of the canonical variables. 
 According to (\ref{feqt}), $T_{abc}=f_{abc}$.  Therefore, the operator of spin connection coefficients is expressed in terms of operators 
 $\hat{f}_{abc}$ found in (\ref{fabcop}): 
 \beq \label{spomegaop}
 \hat{\omega}_{\mu ab} = \half e^c_\mu\left(   \hat{f}_{a bc} + \hat{f}_{ac b} - \hat{f}_{c ab}\right) .
 \eeq
 This expression also depends on the ordering of $e$ and $\der_e$. 
 
 \subsection{The scalar product and the operator ordering} 
 
 The scalar product of Clifford-algebra-valued precanonical wave functions $\Psi(e,x)$  takes the form 
\beq \label{scpr}
\left\langle \Phi | \Psi \right\rangle 
:=  \Tr \int\! {[\rd e]}\ \Phib \,  \Psi , \quad 
\eeq 
where 
 ${[\rd e]}$ is a diffeomorphism invariant measure 
 on fibers of the configuration bundle of tetrads over space-time. 
 Using the arguments similar to those in \cite{misner}
 we obtain 
\beq \label{measure}
{[\rd e]} 
 :=
 {\mathfrak e}{}^{- 4}\prod_{\mu, a} \rd e_\mu^a .
\eeq
 The expectation values of operators are $x-$dependent and calculated according to the formula 
\beq
 \langle \hat{O}\rangle(x) = 
 \Tr \int\! {[\rd e]}\  \Psib (e,x) 
 \hat{O}
 \Psi (e,x) . 
\eeq
However, the norm  $\left\langle \Psi | \Psi \right\rangle $ 
corresponding to (\ref{scpr}) is not positive definite in general. It indicates that a proper subspace of 
positive definite physical wave functions has to be defined by supplementary conditions. An example of such a subspace 
is the projection 
 $\Psi_+:=\frac14(1-\ugamma^0)\Psi(1-\ugamma^0) $ for which 
 $\Tr (\overline{\Psi}_+ \Psi_+)$ coincides with the positive definite Frobenius norm  $\Tr (\Psi_+^\dagger \Psi_+) >0$ if 
 $(\ugamma^0)^2=1$.


Now, that we know the scalar product, we can address the issue of the ordering dependence  of operators obtained in 
(\ref{pmuomop}), (\ref{pmueop}), (\ref{fabcop}), (\ref{hop}) and (\ref{spomegaop}) by requiring them or their densities to be generalized Hermitian in the sense of  (\ref{gherm}).  For example, the generalized Hermitian version of the ordering-dependent operator (\ref{pmuomop}) which satisfies 
\beq
\Tr\!\int {[\rd e]}\  \Psib (e,x) 
 (\hat{\mathfrak{p}}{}_{e^a_\nu} \Psi (e,x)) = 
\Tr\! \int {[\rd e]}\  (\overline{\hat{\mathfrak{p}}{}_{e^a_\nu}\Psi (e,x)}) \Psi (e,x) 
\eeq  
can be easily 
obtained in the form  
\beq
\hat{\mathfrak{p}}{}_{e^a_\nu} := 
 -i\hbar\ka \mathfrak{e} \der_{e^a_\nu} + 
  \frac{3}{2}i\hbar\ka \mathfrak{e} e_a^\nu ,    
\eeq
where $e_a^\nu$ is, obviously, the inverse matrix of 
 $e^a_\nu$:   $e_a^\mu e^a_\nu = \delta^\mu_\nu$.

 
\section{The cosmological constant}

As a consequence of the operator re-ordering in the  operator $\hat{H}$ in (\ref{hop}),  that would make it 
generalized Hermitian with respect to the scalar product (\ref{scpr}) with the measure (\ref{measure}), 
 it  will acquire additional terms proportional to $\sim e\der_e$ and a constant. The constant addition due to re-ordering appears exactly as the cosmological constant if it was included in the original Lagrangian density (\ref{lagrtp}), namely, the extra term $\frac{1}{8\pi G}\mathfrak{e}\Lambda$. It is easy to see that the maximal contribution to the constant emerging from the re-ordering which makes $\hat{H}$ in (\ref{hop}) generalized Hermitian comes from the last term $\frac12 \pi G \hbar^2 \varkappa^2 (e\der_e)^2$ in (\ref{hop}). The numerical value of this constant is easy to estimate: 
$\frac12\times 16^2\times \pi G \hbar^2 \varkappa^2$.  This corresponds to the emergence of the cosmological constant 
\beq \label{llambda}
\Lambda \sim 16 (8\pi G \hbar \varkappa)^2 
\eeq  
which can be interpreted as the quantum-gravitational contribution to the cosmological constant.  
However, this $\Lambda$ depends on the unknown parameter $\varkappa$ introduced by precanonical quantization in the precanonical Schr\"odinger equation 
(\ref{pseq}), in the representation of the operators of polymomenta in (\ref{pmueop}), and in the representation of the classically infinitesimal volume element in (\ref{omegaop}).  

The latter makes it very tempting to identify $\varkappa$ with the Planck scale quantity. In this case, however, the estimation in 
(\ref{llambda}) reproduces “the worst prediction in the history of physics” \cite{sciam}, i.e. the value of the cosmological constant which is 122 orders of magnitude higher than the observed value (see \cite{weinberg,martin} for the reviews). 

However, this identification is not confirmed by our study of precanonically quantized pure Yang-Mills theory where the spectrum of the DW Hamiltonian operator is identified with the spectrum of masses of propagating modes and, therefore, the lowest gap in the spectrum of the DW Hamiltonian operator can be identified with the mass gap $\Delta m$ \cite{my-ymmg}. Our rough estimation for quantum SU(2) gauge theory yields the following result 
\beq  \label{delm}
\Delta m \sim a (g^2\hbar^4 \varkappa)^{1/3} , 
\eeq
where $g$ is the dimensionless (in four dimensions) gauge coupling constant in the classical YM Lagrangian and the constant $ a \approx 0.86$ for SU(2). For  SU(3) it is expected to be a few times greater than for SU(2) because of a higher dimension. 

In the context of QCD, the mass gap is expected to lie between the lowest QCD bound state, which is the neutral pion $\pi^0$ with the mass 135 MeV,  and the glueball candidates at a few GeV \cite{glueballs1,glueballs2}.

In the context of our analysis in \cite{my-ymmg}, the coupling $g$ is the bare coupling in the pure YM Lagrangian. It is not obvious 
which value it takes and how it can be related to the QCD running coupling constant $g_s$ or $\alpha_s := g_s^2/4\pi$ which, in a perturbative renormalization scheme, depend on the renormalization scale. 
Usually, the coupling constant is given at the scale of $M{_Z}\sim$ 91 GeV: $\alpha_S(M_Z)\approx 0.119$,  that gives $g_s (M_Z) \approx 1.2$. The graphs of the running coupling constant show that it changes within an order of magnitude between the deep IR and the far UV scale of the hypothetical Grand Unification at $10^{16}$ GeV. 
 
According to the ``optimized perturbation theory" (OPT) in QCD \cite{coupling1}, which allows us to analyse the infrared properties of the gauge coupling, we can conclude (based on fig. (2b) of \cite{coupling1}) that $g_s(Q=0) \approx 2\sqrt{0.263 \pi} \approx 3.17$. 
If we use the analogy with QED and identify the bare (``classical") coupling constant $g$ in the classical Yang-Mills Lagrangian 
with its deep IR value, we can assume in (\ref{delm})  that $g\approx 3.17$.  
Even if we identify the ``bare coupling constant" $g$ in the Yang-Mills Lagrangian with the value of the running constants of nonabelian theories at the 
GUT scale where the couplings of the weak and strong interactions intersect at the value of $\frac{1}{\alpha_S} (10^{16} \mathrm{GeV}) \approx 1/2$,  we obtain the value of $g_s = \sqrt{4\pi\alpha_s} \approx  5$. 
Thus, for the purpose of the order of magnitude estimations of this  paper, we may assume in (\ref{delm}) that $g^2 \sim 10^1$.  

Consequently, the rough estimation in (\ref{delm}) and the current understanding of the QCD data allows us to identify the scale of 
$\varkappa{}^{1/3}$ with a subnuclear scale 
\beq
\varkappa{}^{1/3} \sim \frac{\Delta m}{ a g^{2/3}} 
\eeq 
(in the units with $\hbar=1$). 
Then,  the cosmological constant estimated in (\ref{llambda})  equals
\beq \label{lambdapre}
\Lambda \sim (32\pi)^2 \ell_{P}^4 \frac{(\Delta m)^6}{a^6 g^4} = 
10^{-46}\ 
\frac{(\Delta m/\mathrm{GeV})^6}{a^6 g^4} \, [\mathrm{cm}^{-2}] , 
\eeq
where  $\ell_{P}:= \sqrt{\hbar G}$ is the Planck length. 
By comparing  (\ref{lambdapre}) with the observed value of the cosmological constant $\Lambda \approx 10^{-56}  \mathrm{cm}^{-2}$  we conclude that 
our formula (\ref{lambdapre}) gives preference to a sub-GeV value of the mass gap $\Delta m$. If we identify 
$\Delta m \sim m_{\pi^0}$ (with a potential error of 1 order of magnitude), and take $g^2 \sim 10$ (with a potential error of 1 order of magnitude) 
and the numerical factor $a \sim 3$ (with a potential error of 1 order of magnitude),  we obtain 
\beq
\Lambda \sim 10^{-46 -5 - 2 -3 \pm (6+1+6)}[\mathrm{cm}^{-2}] = 10^{-56 \pm 13} [\mathrm{cm}^{-2}] 
, 
\eeq
i.e. the observed value of the cosmological constant, albeit with an accumulated error of 13 orders of magnitude 
originating in  
\begin{itemize}  
\item  the roughness of our spectral estimation (\ref{delm}) for SU(2) gauge theory and its extrapolation to SU(3) gauge theory (the undetermined multiplier $a$),    
\item  the uncertainty regarding the value of the mass gap in QCD and its relation to the theoretical mass gap in quantum SU(3) pure gauge theory, 
\item  the uncertainty  regarding the numerical value of the gauge coupling constant $g$ in the classical YM Lagrangian. 
\end{itemize}

Note that similar estimations of the cosmological constant are also possible  in the context of precanonical quantization of vielbein general relativity \cite{ikan1,ikan2,dice22}. The difference is that, in precanonical quantum TEGR,  the cosmological constant emerges from the re-ordering of the DW Hamiltonian operator whereas,  in precanonical quantum GR, the 
 DW Hamiltonian operator is ordering independent and the cosmological constant emerges from the re-ordering of the spin connection term $\sim \what{\gamma}{}^\mu\omega_\mu$ because in this case the curved spacetime Dirac matrices are differential operators with respect to the spin connection coefficients.      

It is also noteworthy that the formula for the cosmological constant which we have obtained using the arguments based on the precanonical quantization of gravity and gauge theory is similar to the formula inferred by Zeldovich in 1967 \cite{zeld67}. The formula connects the cosmological constant with the Planck length $\ell_P$ and the proton mass $m_p$: $\Lambda \sim \ell_P^4 m_p^6$,  and it yields the value of $\Lambda$ eight orders of magnitude higher 
 than the observable cosmological constant, that Zeldovich himself proposed to compensate with a dimensionless coupling of the weak interaction or even a hypothetical weaker interaction. In our formula, instead, the mass gap of SU(3) gauge theory, 
$\Delta m$, replaces $m_p$  and the intuition of Zeldovich regarding the role of a gauge coupling constant is confirmed by the appearance of the factor $g^{-4}$ in (\ref{lambdapre}).

 \section{Conclusion} 

 We have shown that the scheme of precanonical quantization of GR (in vielbein Palatini formulation)  can be extended to TEGR (in tetrad Palatini formulation). The structure of constraints of the DW Hamiltonian formulation and the resulting Dirac brackets, and the reduction of the polymomentum phase space are more complicated for TEGR. On the quantum level, TEGR is described in terms of the Clifford-algebra-valued wave function on the frame bundle over space-time: $\Psi(e_\mu^a, x^\nu)$, whereas the precanonical quantization of GR naturally leads to the description in terms of the wave function on the spin connection bundle: $\Psi(\omega_\mu^{IJ}, x^\nu)$. There is one physical property of precanonical quantum GR which does not have an obvious teleparallel analog: the quantum avoidance of the curvature singularity in the former due to the normalizability requirement for the precanonical wave function. Another difference is that the precanonical Schr\"odinger equation for quantum GR decouples odd and even parts of the Clifford-valued precanonical wave function, whereas the precanonical Schr\"odinger equation for quantum TEGR can be closed on the subspace of scalar and vector parts (in the language of Clifford algebras)  of the precanonical wave function like it is in the case for precanonically quantized scalar fields on Minkowski space-time \cite{ik5e}. 
 
 Both quantum TEGR and quantum GR lead to very similar estimations of the cosmological constant based on the arguments related to the proper ordering of operators in agreement  with the respective scalar products.  The critical part of the argument is our previous spectral estimate for the gap in the spectrum of the DW Hamiltonian operator of a nonabelian gauge theory \cite{my-ymmg} and its interpretation as the mass gap in QCD phenomenology and lattice gauge theories. These considerations allow us to claim that the scale of the parameter $\varkappa$ in precanonical quantization is subnuclear, making the value of the cosmological constant derived from the considerations based on precanonical quantization very close to the observable value of $\Lambda$.  Unfortunately, the related uncertainties   currently lead to the estimation of the accumulated error  of our theoretical prediction of the value of $\Lambda$ of 13 orders of magnitude.  This error, however, can be improved in the future in many ways, starting from an improved estimation of the spectrum of the DW Hamiltonian for pure SU(3) Yang-Mills theory to the consideration of the spectrum of QCD bound states like pions within the precanonically quantized full QCD, which are expected to lead to a more precise estimation of the value of $\varkappa$ and, therefore, a more precise estimation of $\Lambda$.

\subsection*{Acknowledgement} I thank M.E. Pietrzyk, C. Barbachoux, J. Kouneiher and V.A. Kholodnyi for their interest, 
support and collaboration.


\end{document}